\documentclass[aps,pra,twocolumn,showpacs,amsfonts,amssymb,amsmath]{revtex4}  

\usepackage{graphicx}

\newcommand{\EXP}[1]{\mathrm{e}^{#1}} 

\newcommand{\DEFt}{\overset{\text{\tiny def}}{=}}

\newcommand{\ket}[1]{|#1\rangle} 
\newcommand{\kete}[1]{|\kern.3ex#1\kern.3ex\rangle}

\newcommand{\brae}[1]{\langle\kern.3ex #1 \kern.3ex|} 
\begin{document}

\title{Influence of classical resonances on chaotic tunnelling}

\author{Amaury Mouchet}
\affiliation{$^1$Laboratoire de Math\'ematiques 
  et de Physique Th\'eorique 
  \textsc{(cnrs umr 6083)}, Universit\'e Fran\c{c}ois Rabelais
  Ave\-nue Monge, Parc de Grandmont 37200
  Tours,  France.}

\email[email: ]{mouchet@phys.univ-tours.fr}
\author{Christopher Eltschka}
\author{Peter Schlagheck}
\affiliation{Institut f\"ur Theoretische Physik,
Universit\"at Regensburg,
93040 Regensburg, Germany
} 
\email[email: ]{Peter.Schlagheck@physik.uni-regensburg.de}
\date{\today}


\begin{abstract}

Dynamical tunnelling between symmetry-related stable modes is studied in the 
periodically driven pendulum.
We present strong evidence that the tunnelling process is governed by nonlinear
resonances that manifest within the regular phase-space islands on which the
stable modes are localized.
By means of a quantitative numerical study of the corresponding Floquet
problem, we identify the trace of such resonances not only in the level
splittings between near-degenerate quantum states, where they lead to 
prominent plateau structures, but also in overlap matrix elements of the
Floquet eigenstates, which reveal characteristic sequences of avoided
crossings in the Floquet spectrum.
The semiclassical theory of resonance-assisted tunnelling yields good overall
agreement with the quantum-tunnelling rates, and indicates that partial
barriers within the chaos might play a prominent role.

\end{abstract}

\maketitle

\section{Introduction}

Since the very first quantitative studies of tunnelling in a chaotic
system~\cite{Lin/Ballentine90a,Grossmann+91a}, it has been clear that
a minute scrutiny of the associated classical dynamics was required in
order to understand even the most coarse features of the quantum
behaviour of such systems.  In essence, tunnelling is a semiclassical
concept since it refers to a quantum process --- typically a decay or
the oscillation of an averaged observable --- that is forbidden at a
classical level. But, despite the numerous successes of semiclassical
computations in quantum chaos, the questions of which and how
classical objects can be used to understand tunnelling and to compute,
say, their characteristic time scales, have been remaining widely open
for sixteen years.  The stakes in the battle are important since
non-integrability is the generic rule of multidimensional systems and
tunnelling may play a crucial r\^ole in their transport
properties. Moreover, it is one of its signatures that chaotic tunnelling can
be modified on several orders of magnitude by the slightest variation
of any classical or quantum parameter; therefore a deep understanding
of chaotic tunnelling is required to control
the process, what may be an advantage
in delicate quantum experiments and, hopefully, give rise to an extremely
sensitive quantum tool. Some promising clues have been provided in
this direction by numerical studies and experiments with cold
atoms~\cite{Mouchet+01a,Hensinger+01a,Steck+01a,Hensinger+04a} but
also with microwave cavities~\cite{Dembowski+00a,Hofferbert+05a} where
tunnelling signatures for processes that are forbidden by ray optics were
observed.

Of course, in order to capture the typical exponentially small
tunnelling effects, it is expected that the classical dynamics should
be complexified.  It is
well-known~\cite{Landau/Lifshitz58c,Balian/Bloch74a} that the complex
solutions of Hamilton's equations are actually involved in the
interpretation of tunnelling of autonomous systems with one degree of
freedom.  In the mid 90's, the first
observations~\cite{Kus+93a,Leboeuf/Mouchet94a,Scharf/Sundaram94a,Creagh/Whelan96a} that
complex periodic orbits allow one to reproduce quantitatively some feature
of chaotic tunnelling gave hope that a semiclassical strategy was
indeed possible, even though the complexified classical tori are
generically destroyed in chaotic systems \cite{footnote}.
But to deal with tractable semiclassical trace formulae \`a la
Gutzwiller, a general criterion for selecting the complex
periodic orbits was still lacking; this need became an emergency when
it was unexpectedly discovered~\cite{Shudo/Ikeda95a,Shudo/Ikeda98a} 
that chaos reveals itself in the 
complex phase-space through some fractal structures, the so-called Laputa
islands, that look like agglomerates of complex trajectories.  It is
only recently that some encouraging significant steps were done
for retaining the relevant semiclassical skeleton~\cite{Shudo+02a}. A
lot of work remains to be done in that direction, especially 
if one wants to deal with continuous systems where time can (and must) 
be complexified as well, unlike what occurs in discrete maps.

The second strategy to cope with chaotic tunnelling is not purely
semiclassical but rather calls up random matrix theory.  Since
the seminal work presented in Ref.~\cite{Bohigas++83A}, a fruitful approach of
quantum chaos is to replace a chaotic but deterministic Hamiltonian by a
random element of an ensemble of matrices that only 
encapsulates the global symmetries. These hybrid techniques, with both
semiclassical and statistical ingredients, first allowed us to 
qualitatively understand the so-called chaos-assisted tunnelling, \textit{i.e}
the observation~\cite{Bohigas+93a} that tunnelling is increased \emph{on
average} as the transport through chaotic regions
grows~\cite{Bohigas+93b,Tomsovic/Ullmo94a,Leyvraz/Ullmo96a}.  However,
the extreme sensitivity of tunnelling renders the predictions very
difficult even if just an order of magnitude is required. 
This is also true for the seemingly simple case of \emph{near-integrable}
dynamics where it was shown, on a discrete quasi-integrable quantum map,
that the internal resonances may enhance the transitions by several orders of
magnitudes~\cite{Brodier+01a,Brodier+02a}.  This resonance-assisted
tunnelling is also at work in discrete systems where chaos is much
more developed~\cite{Eltschka/Schlagheck05a}. The aim of the present
paper is to show that the ideas in Refs.~\cite{Eltschka/Schlagheck05a,Schlagheck+05a} are
strengthened, now in a continuous system, by a systematic analysis of
level dynamics and the phase-space representation of the quantum states.

We shall begin in section~\ref{sec:model} with a short presentation of
the general framework of chaotic tunnelling and the model we choose in
order to study it. We will be concerned with a typical 
signature of tunnelling, namely the period of Rabi oscillations between two
wells that are separated by a dynamical barrier.
Then we will show (section~\ref{sec:internalstructure}) with a simple
argument, that the attempt to reproduce the average tunnelling periods
presented in Ref.~\cite{Podolskiy/Narimanov03a} is far from being complete
precisely because it ignores the resonances, among other things. In
section~\ref{sec:multilevels} we 
will give an illuminating illustration of a characteristic feature of
chaotic tunnelling~\cite{Mouchet/Delande03a} : it appears to be a
collective effect in level dynamics where not just one third state
crosses a tunnelling doublet.  We will give a
phase-space picture of resonance-assisted tunnelling and confirm,
 that
taking into account the resonances is unavoidable if we want to
reproduce or predict the average behaviour of tunnelling transitions. 
In section~\ref{sec:rat}, we show that the ideas of
Refs.~\cite{Eltschka/Schlagheck05a,Schlagheck+05a} actually provide a good
estimate for the average tunnelling rate in our model.

\section{General framework of chaotic tunnelling}\label{sec:model}

The simplest non-integrable Hamiltonian models are either
time-dependent one-dimensional (1D) systems or, equivalently,
autonomous systems with two degrees of freedom where the Hamiltonian
is the only constant of motion. Seen from the classical point of view,
 a generic potential induces a cascade of non-linear
resonances whose overlap generates
chaos~\cite{Chirikov79a,Lichtenberg/Lieberman83a}.  One minimal
continuous model that encapsulates these typical properties is a
1D time-dependent system whose Hamiltonian is
\begin{equation}\label{eq:ham}
  H(p,q;t)
  =\frac{p^2}{2} -\frac{\gamma_+}{2} \ \cos(q+2\pi t/\tau) 
  -\frac{\gamma_-}{2} \ \cos(q-2\pi t/ \tau)\;,
\end{equation}
where $p$ and~$q$ denote canonical action-angle variables that
are coordinates on a phase-space having the topology of a cylinder:
it is $2\pi$-periodic in position (angle)~$q$ and infinitely extended in
the momentum (action) direction. The model~\eqref{eq:ham} can be
seen as the most natural normal form where we keep only two overlapping
resonances \textit{i.e.} the first two time-dependent Fourier
components of the potential. The Hamiltonian~$H$ can also be
interpreted, provided a change of frame is performed, as a pendulum
driven by a periodic wave.  Moreover, it is actually very similar to
the effective Hamiltonian that can be realised in experiments on cold
atoms \cite{footnote2}.
We shall consider the period $\tau \equiv 2 \pi$ in the following.

\begin{figure}[!ht]
\center
\includegraphics[width=8.5cm]{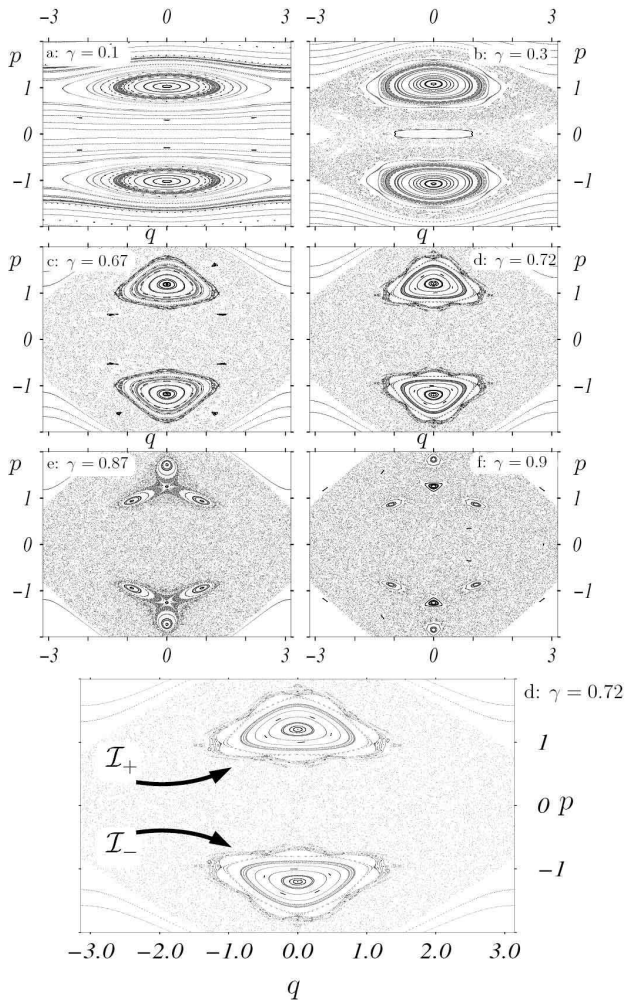}
\caption{\label{fig:2yeux} Some Poincar\'e sections
($\tau=2\pi$-stroboscopic plots) corresponding to
Hamiltonian~\eqref{eq:ham} for several values
of~$\gamma_+=\gamma_-=\gamma$.  Increasing~$\gamma$, between
$\gamma\simeq0.1$ and~$\gamma\simeq0.3$, the two resonances
at~$(p,q)\simeq(\pm1,0)$ start to significantly overlap,
and the systems enter in a mixed regime where both regular and chaotic
dynamics coexist. From $\gamma\sim0.3$ to $\gamma\sim0.8$, the two
stable islands~$\mathcal{I}_\pm$ are completely surrounded by a
chaotic sea, and although they are related by the time-reversal
symmetry, they are disconnected from each other at a classical
level, since dynamical barriers prevent
the real classical trajectories
starting in one island to escape from
it. For~$\gamma\sim0.72$,~$\mathcal{I}_\pm$ have developed a
relatively wide $3/7$ resonance,
which is not the case for~$\gamma\sim0.67$ where the
corresponding $3/7$~resonance chain is hardly visible. The insets in
FIG.~\ref{fig:comparison} show magnifications of the islands in 
these last two cases.}
\end{figure}

The two parameters~$\gamma_\mp$ control the size of the two stable
islands~$\mathcal{I}_\pm$ located in phase-space near~$p=\pm1$
respectively. When increasing the~$\gamma$'s from zero, we leave the
quasi-integrable free-like motion and rapidly (at~$\gamma\sim0.2$)
reach a mixed regime where the two stable resonance islands are fully
surrounded by a chaotic sea. If an initial condition lies inside one
island, the classical motion will remain trapped forever 
within a very thin quasi-one dimensional layer without the possibility 
of escape (this evasion is the forbidden process that 
tunnelling will be concerned
with~\cite{Davis/Heller81a}).  Alternatively, a trajectory starting in
between the two islands is chaotic: without being fully ergodic since
it cannot enter the islands, the absence of a sufficiently large
number of constants of motion allows the system to explore large areas
and to develop an exponential sensitivity on initial conditions.
Still increasing the~$\gamma$'s, the cascade of secondary resonances
inside the islands erode gradually the regular zones and make them
dissolve completely in the chaotic sea. Figure~\ref{fig:2yeux} shows some 
Poincar\'e sections of the classical dynamics,
\textit{i.e} views taken stroboscopically at every integer multiple of
the time period~$\tau=2\pi$ of~\eqref{eq:ham}.

At a quantum level, the quantization of a Hamiltonian
like~\eqref{eq:ham} has, for a long time, been employed as
a very natural continuous model for studying 
tunnelling~\cite{Latka+94a,Averbukh+95a,Bonci+98a}. The time-dependence is
implemented within the Floquet theory where the phases~$\epsilon$ of
the eigenvalues of the evolution operator at time $\tau$ play the
r\^ole of the energies for autonomous systems.  The eigenvectors
associated with these so-called quasi-energies $\epsilon$ will be
called the (eigen)states of the system. When~$\hbar$ is small enough 
compared to the typical size of the stable islands, some states, 
the \emph{regular} states~$\{\ket{\phi_n}\}_{n\in\{0,1,\dots\}}$, 
appear to have their Husimi distribution localised on the two main  stable 
islands while other states, the \emph{chaotic}
states, are found to be delocalised in the chaotic sea.

Following the same route that is explained in great detail
in Ref.~\cite[\S\S~II \& III]{Mouchet/Delande03a}, we will
keep a discrete two-fold symmetry, namely the time-reversal symmetry,
by considering~$\gamma_+=\gamma_-=\gamma$ throughout this paper~\cite{footnote3}.
This allows us to clearly identify the tunnelling process through the 
existence of small energy scales, the doublet splittings~$\Delta\epsilon_n$, 
which are associated with large time scales, namely
the periods~$2\pi\hbar/\Delta\epsilon_n\gg\tau$ of tunnelling oscillations
between the stable islands. In other words, the magnitude of the 
splittings~$\Delta\epsilon_n=|\epsilon_n^+-\epsilon_n^-|$
 between the states that are symmetric ($\ket{\phi_n^+}$) and antisymmetric 
($\ket{\phi_n^-}$) with respect
to the time-reversal symmetry measure the importance of tunnelling between  
the two stable islands.  The actual challenge of chaotic tunnelling is to 
understand and hopefully predict the behaviour of~$\Delta\epsilon_n$ as 
a classical~($\gamma$) or quantum ($\hbar$) parameter is varied within a regime where
the quantum scales are small enough to resolve the classical scales of 
the chaotic structures. 

Unlike what occurs for integrable multidimensional systems where
$\Delta\epsilon_n$ is a smooth monotonic function given by
$\Delta\epsilon\propto\EXP{-A/\hbar}$~ \cite{Creagh94a} with $A$ being a
typical classical action that characterizes the tunnelling barriers,
the splittings can display huge fluctuations in the mixed
regular-chaotic case, which were soon identified as a signature of
chaos~\cite{Tomsovic/Ullmo94a,Latka+94a}.  But as far as
\emph{isolated} fluctuations are observed this point of view must be
amended.  One isolated fluctuation is clearly associated with a third
state whose (quasi)energy nearly degenerates with the
doublet~\cite{Latka+94a,Bonci+98a} as $\gamma$ or $\hbar$ is
varied. Chaos is not necessarily relevant here since such fluctuations
can be observed when the chaotic layers are too small (and~$\hbar$ too
large) to be resolved by the quantum waves. It may happen that the
third state is also a regular state localised on  another regular
\textsc{ebk} torus, possibly belonging to another stable island. What
deserves the name of chaotic tunnelling is a radical change of regime
where the fluctuations are not isolated anymore and where the
coarse-grained behaviour of~$\Delta\epsilon$ does not follow a
monotonic law. It was shown in Ref.~\cite[\S~V]{Mouchet/Delande03a},
that there actually exists a rather
abrupt transition between a quasi-integrable tunnelling regime and a
chaotic regime and that this transition occurs precisely when the
quantum eyes can resolve the chaotic classical structures.

While the study of statistics of the splittings is necessary to give an
insight into chaotic
tunnelling~\cite{Leyvraz/Ullmo96a,EgydiodeCarvalho/Mijolaro04a},
we will focus here on the \emph{average} behaviour of the
splittings. It is far from obvious that we can conceptually
justify the distinction between ``large'' scales (the average) and 
``small'' scales (the fluctuations) in the variations of~$\Delta\epsilon$. 
Most probably, there may exist a whole hierarchy
of such variations. From the numerical point of view, however,
there seems to be an overall modulation of~$\Delta\epsilon$, both observed
in maps (\cite[FIG.~3]{Roncaglia+94a}, \cite[FIGs.~1
\&~2]{Eltschka/Schlagheck05a}) and in continuous
systems~\cite[FIG.~7]{Mouchet+01a}. This modulation is precisely the
very object of Ref.~\cite{Eltschka/Schlagheck05a} (see also
Ref.~\cite{Schlagheck+05a}) and of the present paper.

\section{Key r\^ole of resonances}\label{sec:internalstructure}

In the general context recalled in the previous section, where tunnelling 
transitions occur between two symmetric but disconnected stable classical
islands, the following estimate has been proposed for a typical chaotic 
tunnelling splitting~\cite[eq. (4)]{Podolskiy/Narimanov03a}:
\begin{equation}\label{eq:PNestimate}
  \Delta\epsilon\simeq \hbar \Omega \frac{\Gamma(2N,4N)}{\Gamma(2N+1;0)} \, .
\end{equation}
Here, $\Gamma$ stands for the incomplete Gamma function, 
$N =A/(2\pi\hbar)$ denotes the semiclassical estimate of the number of
states localised in one island of area~$A$, and $\Omega$ represents an
unknown prefactor, with the dimension of an inverse time scale, which
does not depend on $\hbar$.  Though the origin of the formula remains
obscure as Eq.~\eqref{eq:PNestimate} is not explicitly proven by their
authors, it gave, in Ref.~\cite{Podolskiy/Narimanov03a}, good
agreement with numerical computations provided we are ready to accept
an unreasonably large ambiguity on the unspecified proportionality
factor: for~$\hbar^{-1}\simeq40$, the estimate~\eqref{eq:PNestimate}
varies by five orders of magnitude ($10^{-1}$ versus~$10^{-6}$) in the
two cases considered in FIG.~2 of Ref.~\cite{Podolskiy/Narimanov03a}
where the areas~$A$ of the stable islands are of the same order.  More
recent calculations within the ``kicked Harper'' model, however,
revealed substantial deviations between Eq.~\eqref{eq:PNestimate} and
the exact quantum tunnelling rates in the semiclassical regime, where a
reasonable agreement was only found in the deep quantum limit of large
$\hbar$ \cite{Schlagheck+05a}.

\begin{figure}[!ht]
\center
\includegraphics[width=7.5cm]{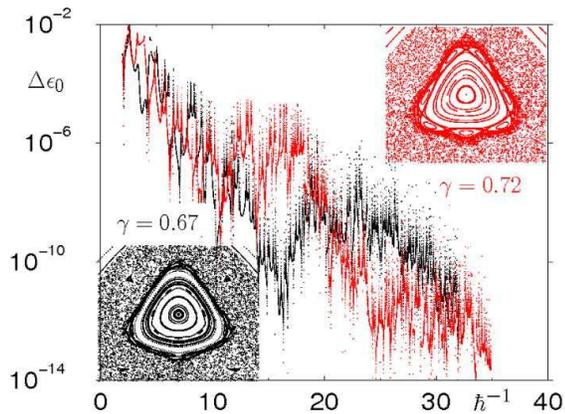}
\caption{\label{fig:comparison}
(Color online) Comparison of the
tunnelling splittings~$\Delta\epsilon_0$ for~$\gamma=0.72$ (black
line) and~$\gamma=0.67$ (gray/red line). The corresponding
stable islands, magnified in both
cases with the same scaling factor,
are shown in the insets
(see FIG.~\ref{fig:2yeux}~c) and~d)
for a full Poincar\'e surface of section).  The 
exponentially large
difference between the two curves is explained by the different
internal structure of the islands: for~$\gamma=0.72$, the $3/7$
resonance chain is much more developed than for~$\gamma=0.67$.}
\end{figure}

Indeed, no good estimate of~$\Delta\epsilon$ can be obtained if the only classical
parameter on which the theory depends is the area~$A$ of the island. As shown
in FIG.~\ref{fig:comparison},
the internal classical structure of the islands must be involved 
in one way or another. We have plotted here the
splitting~$\Delta\epsilon_0$ between the two ``central''
states~$\ket{\phi_0^\pm}$ localised in the two islands; more 
precisely, they both were
selected by the criterion of having the maximal
overlap with a coherent state that is
located on the central stable periodic
orbit of period~$\tau$. The two graphs, $\Delta\epsilon_0$ as a
function of~$1/\hbar$, are shown for 
the two classical parameters~$\gamma=0.67$ and~$\gamma=0.72$ (see also
FIG.\ref{fig:2yeux}, c) and d)).  In these 
cases, the stable symmetric
islands~$\mathcal{I}_\pm$ have an area~$A$ of the same order of
magnitude, but exhibit
a rather distinct internal structure: for~$\gamma=0.72$,
$\mathcal{I}_\pm$ have developed a wide $3/7$-resonance chain compared
to the~$\gamma=0.67$ case. In the
log-plot, the two graphs strongly
differ.  Not only the fluctuations of~$\Delta\epsilon_0$ hardly match,
but also the average behaviour is
completely distinct in a semiclassical
regime where~$1/\hbar>10$.  A discrepancy of about five orders of
magnitude can be clearly observed for $1/\hbar\simeq16$:
for~$\gamma=0.72$, we have~$\Delta\epsilon_0\sim10^{-6}$ compared
to~$\Delta\epsilon_0\sim10^{-11}$ for~$\gamma=0.67$.  At
$1/\hbar\sim25$, on the other hand, the splittings for $\gamma=0.72$
are about $10^{-3}$ smaller than for $\gamma=0.67$.

In any case, no estimate formula~$\Delta\epsilon_0(A,\hbar)$, which would 
produce approximately the same graphs for~$\gamma=0.67$ and
$\gamma=0.72$, can be satisfactory. At this stage it is more than
plausible that the internal structure of~$\mathcal{I}_\pm$ must be
taken into account in any attempt to estimate~$\Delta\epsilon_n$ with
the help of classical ingredients.

\section{Multilevel crossing}
\label{sec:multilevels}

To obtain more insight, we analyze in this section the ``level
dynamics'', \textit{i.e.}  the changes of the quasi-energy spectrum as
$1/\hbar$ or~$\gamma$ is varied.  This can, for instance, explain individual
fluctuations spikes, which correspond to the crossing of the tunnelling
doublet with a third state (or an unresolved
doublet)~\cite[FIG.~7]{Tomsovic/Ullmo94a}
\cite[FIG.~1]{Latka+94a},\cite[FIG.~4]{Mouchet/Delande03a}. As far as
the average of~$\Delta\epsilon$ is concerned, we can also establish a
correspondence between its behaviour as a function of $1 / \hbar$ and
some features of the level dynamics. First, a technical point should be
mentioned at this stage.  For $\tau$-periodic systems whose classical
phase-space is unbounded, it is expected that the quasi-energy
spectrum becomes a dense set because of the foldings of an infinite
spectrum in the finite Floquet
zone~$]-\pi\hbar/\tau,\pi\hbar/\tau]$. Hence, we need a criterion to select only those
states that are relevant in the level dynamics.
Numerically the truncation of the Floquet matrices at high $|p|$'s is
not sufficient, especially for small~$\hbar$, since more and more
regular and chaotic states are \textit{a priori} potentially
implicated in the tunnelling dynamics.  To select the levels that are
actually involved, we will use a simple criterion based on the
systematic computation of the overlaps~$\sigma_m$ between the
states~$\ket{\psi_m}$ and a coherent state centered on the island:
$\sigma_m\DEFt|\langle z_{\mathrm{stable\ orbit}}\ket{\psi_m}|$.
Retaining only those states whose~$\sigma_m$ is larger than a given
threshold~$\sigma_{\mathrm{filter}}$ will filter the levels that are
the best candidates to play a r\^ole in the tunnelling process, precisely
because their wave-functions are not negligible in the area where the
tunnelling doublet wave-functions live.

\begin{widetext}\center
\begin{figure}[!ht]
\center
\includegraphics[height=18cm, angle=-90]{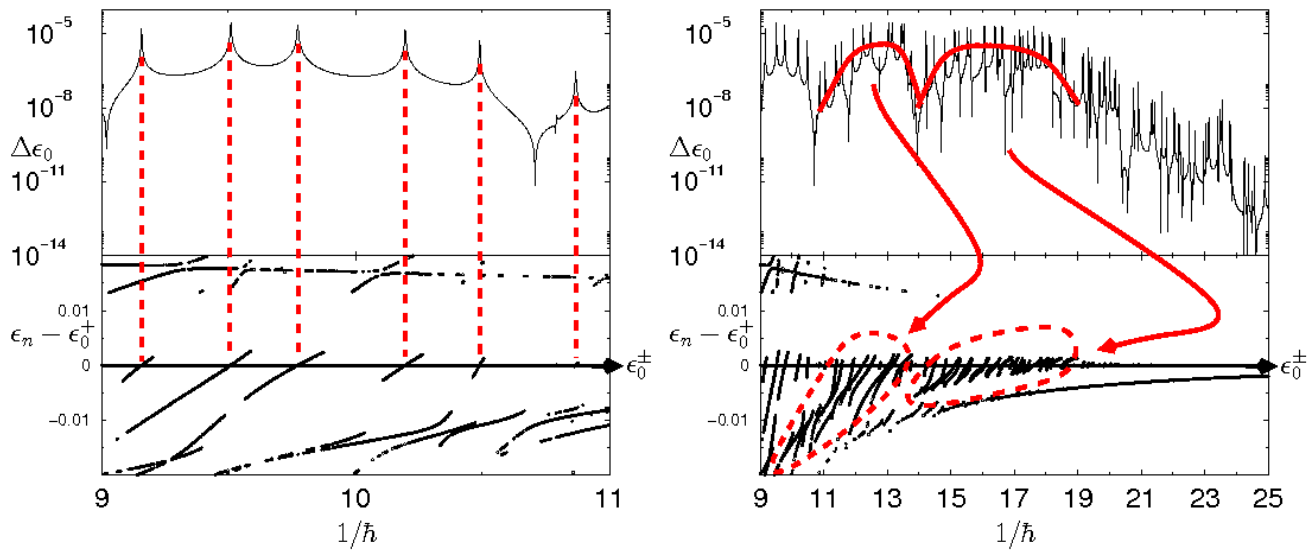}
\caption{\label{fig:multilevel}
(Color online) The upper panels show $\Delta\epsilon_0$ as a function of~$1/\hbar$
for~$\gamma=0.72$ and the lower panels display a part of the
corresponding Floquet spectrum.  Only the quasi-energies associated
with states whose overlaps are~$\sigma_m\geq7\times 10^{-3}$ are shown, once
calibrated with respect to the tunnelling doublet~$\epsilon_0^\pm$
(unresolved in the lower panels).  For relatively large $\hbar$ (on
the left-hand side), the huge fluctuations of $\Delta\epsilon_0$ are in
one-to-one correspondance with the crossings of the tunnelling doublet
by one isolated doublet. On the right-hand side, the upper
panel shows a magnification of the plateau in~$\Delta\epsilon_0$
occurring at $\gamma=0.72$ between~$\hbar\simeq11$ and~$\hbar\simeq19$
(see also FIG.~\ref{fig:comparison}).  The lower right panel displays
a part of the corresponding Floquet spectrum.  The plateau
in~$\Delta\epsilon_0$ corresponds to a large number of crossings
where we can even identify by eye two families of levels 
(encircled by the dashed lines) that give rise
to the two bumps in the average of~$\Delta\epsilon_0$
(red/grey thick continuous line in the upper right panel).
Beyond the point where the plateau ends, i.e.\ for~$\hbar^{-1}\gtrsim19$, no more
crossings can be identified, and the only other state apart from $\epsilon_0^\pm$ that
significantly overlaps with the Husimi function in the island is the bystander
doublet~$\epsilon_5^\pm$ (see also FIG.~\ref{fig:ratl7}).
  }
\end{figure}
\end{widetext}

Let us fix the classical dynamics at~$\gamma=0.72$ where, as we have
seen in the previous section, the internal $3/7$-resonance is
suspected to provoke the enhancement of average tunnelling for
$11\lesssim1/\hbar\lesssim19$ (see FIG.~\ref{fig:comparison}).
When we look at this plateau more carefully, we can identify two
``bumps'' that are visualised in FIG.~\ref{fig:multilevel}. One
corresponds to~$11\lesssim1/\hbar\lesssim13.5$ followed by a wider
one for~$13.5\lesssim1/\hbar\lesssim19$. These two bumps are in
one-to-one correspondence with two bunches of levels with significant
overlaps~$\sigma_m$ crossing the tunnelling
doublet~$\epsilon_0^\pm$. More generally, we have observed in many cases,
with such a level-dynamics point of view, that there seems
to be a clear change of regime: for a
given~$\sigma_{\mathrm{filter}}$, when the average
of~$\Delta\epsilon_n$ stops decreasing,
the states~$\ket{\psi_m}$ whose
quasi-energies are in the neighbourhood of~$\epsilon_n^\pm$ and
whose~$\sigma_m$ are larger than $\sigma_{\mathrm{filter}}$
become significantly
more numerous.  For instance, in FIG.~\ref{fig:multilevel} where we
chose~$\sigma_{\mathrm{filter}}\simeq7\times 10^{-3}$, or in
FIG.~\ref{fig:ratl7} with a five times more rough
filter~$\sigma_{\mathrm{filter}}\simeq1.4\times 10^{-3}$, just after the
end of the large plateau at~$1/\hbar\gtrsim20$, the crossings with
relevant levels become suddenly scarce. To put it in a different way,
the average enhancement of tunnelling appears to be the outcome of a
collective dynamics involving numerous states that can be seen in the
spectrum through their crossing with the tunnelling doublet.

\begin{widetext}\center
\begin{figure}[!ht]
\center
\includegraphics[width=13cm]{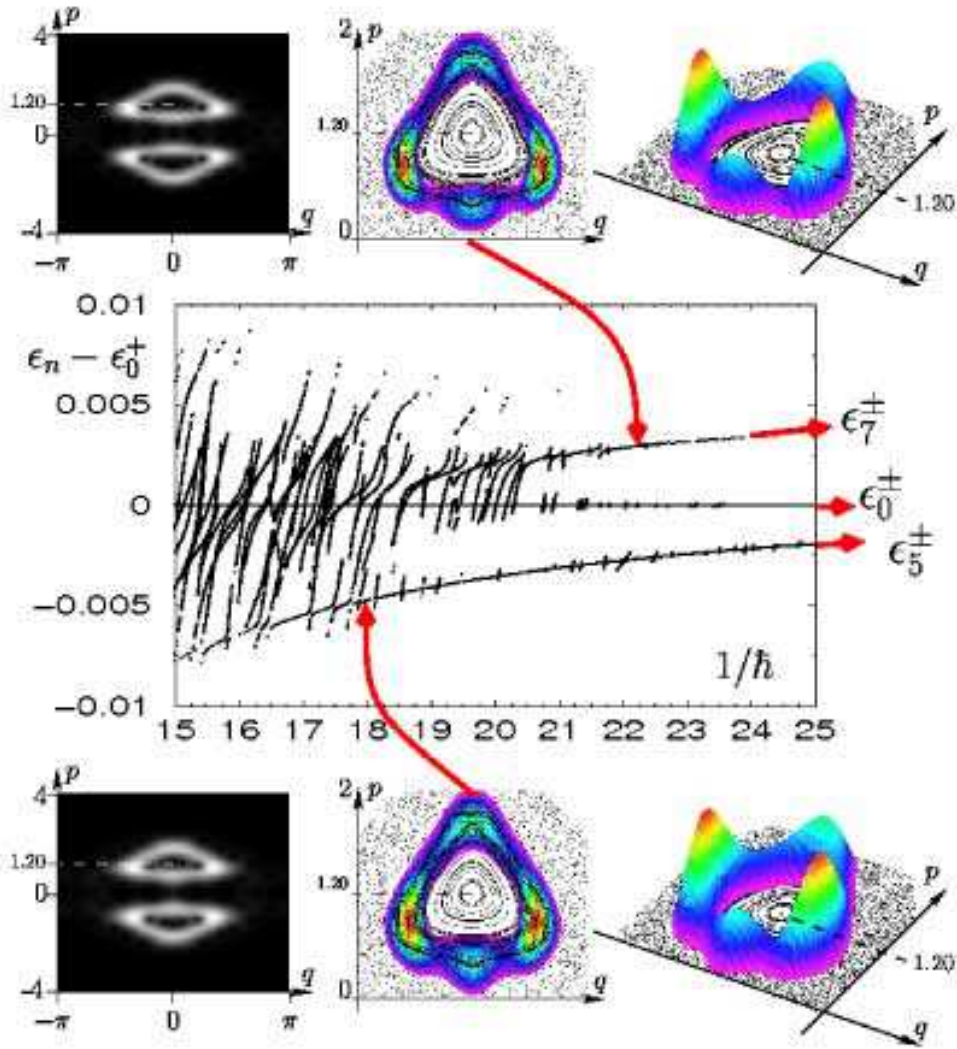}
\caption{\label{fig:ratl7}
(Color online) The level dynamics for~$\gamma=0.72$
 calibrated with respect to the tunnelling doublet~$\epsilon_0^\pm$
(unresolved).  Compared to FIG.~\ref{fig:multilevel} more states are
 plotted since we take a smaller~$\sigma_{\mathrm{filter}}\simeq1.4\times 10^{-3}$. 
When looking at the crossings with the tunnelling doublet,
 there still is a clear transition of regime for~$\hbar^{-1}$ below or
 above~$20$, which corresponds
 to the end of the plateau
 in~$\Delta\epsilon_0$. The three upper views show the Husimi
 distribution of the state, $\ket{\phi_7^+}$,
 whose quasi-energy is~$\epsilon_7^+$, for~$\hbar^{-1}>20$.
 A comparison with the Poincar\'e  surface of section in the
 neighbourhood of the stable island~$\mathcal{I}_+$ is
 shown. The three lower views
(which are, actually, 
very similar to the upper ones at this resolution)
 correspond to $\ket{\phi_5^+}$,
the symmetric state
whose quasi-energy is~$\epsilon_5^+$ can be followed for every~$\hbar^{-1}$.
Unlike $\ket{\phi_7^\pm}$, the doublet  $\ket{\phi_5^\pm}$ does not cross and 
has no influence on the tunnelling doublet. 
$\ket{\phi_7^+}$ (resp. $\ket{\phi_5^+}$) is localised near
the boundary of the two islands and indeed corresponds to
the symmetric combination of the 7th
 (resp. 5th) excited regular state in~$\mathcal{I}_{\pm}$. 
}
\end{figure}
\end{widetext}

In order to be more convincing, this qualitative observation can be
strengthened by a phase-space analysis.  In
addition to the unresolved tunnelling doublet~$\epsilon^0_\pm$, we
can clearly see, in the level dynamics in FIG.~\ref{fig:ratl7},
 two other unresolved doublets~$\epsilon_5^\pm$
and~$\epsilon_7^\pm$ for~$1/\hbar\gtrsim20$ whose Husimi distribution is mainly
located near the boundary of the stable
islands~$\mathcal{I_\pm}$. In fact, the doublet~$\epsilon_5^\pm$ can
be followed along a quasi-continuous line even for~$1/\hbar\leqslant20$
whereas the other one cannot be identified unambiguously in that
region of~$\hbar$ corresponding to the plateau: when~$1/\hbar$ is
decreased from about~$19$, the doublet line~$\epsilon_7^\pm$ encounters
many avoided crossings and ramifies into the bunch of levels we have
precisely associated with the bump.  Therefore, even if their Husimi
plots both look very similar, one doublet is dramatically involved in
the tunnelling process while the other remains a bystander.  If we
compute the overlap of these states with the harmonic states inside
the islands, that is, the eigenstates of the harmonic approximation of
the Hamiltonian~\eqref{eq:ham} near the corresponding stable periodic
orbits, we find that the bystander doublet~$\epsilon_5^\pm$ is indeed
the 5th excited doublet in~$\mathcal{I}_\pm$ whereas the
other~$\epsilon_7^\pm$ is the~7th. 
This can also be checked in the Floquet spectrum. Indeed, we expect that the
 levels of the local eigenmodes of the island (which is locally equivalent to
 a harmonic oscillator) approximately differ from each other by multiples of
 $\hbar \omega_0$ where $\omega_0 \simeq 0.4$ denotes the frequency
 oscillations
 around the center of the island at~$(p,q)\simeq(\pm1.20,0)$. This yields
 $\epsilon_l^\pm\simeq\epsilon_0^\pm+l\hbar\omega_0$ modulo a Floquet
 width~$2\pi\hbar/\tau$ for the levels. 
 The fact that it is precisely the quantum
number~$\ell=7$ which is involved in the emergence of the plateau is
not a coincidence. This~$\ell$ is exactly the order of the
resonance~$3/7$ that dominates in~$\mathcal{I}_\pm$ for~$\gamma=0.72$.

The quantitative details of how the classical resonances may be
implemented in order to reproduce the average behaviour
of~$\Delta\epsilon_0$ will be described in the next section. Even though
the approximations that are involved are not always under rigorous
control, we can see here the resonances at work. Suppose
that the classical parameters are such that one \emph{classical}
resonance~$s/\ell$ dominates the others in the stable
islands~$\mathcal{I}_\pm$ ($s$ being an integer and $\ell$ a strictly
positive integer, the order of the resonance).  This precisely means
that one torus is actually broken into a chain of~$\ell$ sub-islands
centered about a stable periodic orbit of
period~$2\pi/\omega_0=\ell\tau/s$. A \emph{quantum} resonance occurs
when two quasi-energies~$\epsilon_m$ and~$\epsilon_0$ are nearly degenerate, 
\textit{i.e.} differ by an integer number~$s'$ of Floquet
widths: $\epsilon_m-\epsilon_0\simeq s'\hbar2\pi/\tau$. As mentioned above, we
also have $\epsilon_m-\epsilon_0\simeq
m\hbar\omega_0+s''2\pi/\tau$ from a semiclassical argument ($s''$ being an integer). 
Hence, we immediately see that~$m$ must be an integer multiple of~$\ell$.

\begin{figure}[!ht]
\center
\includegraphics[width=8.5cm]{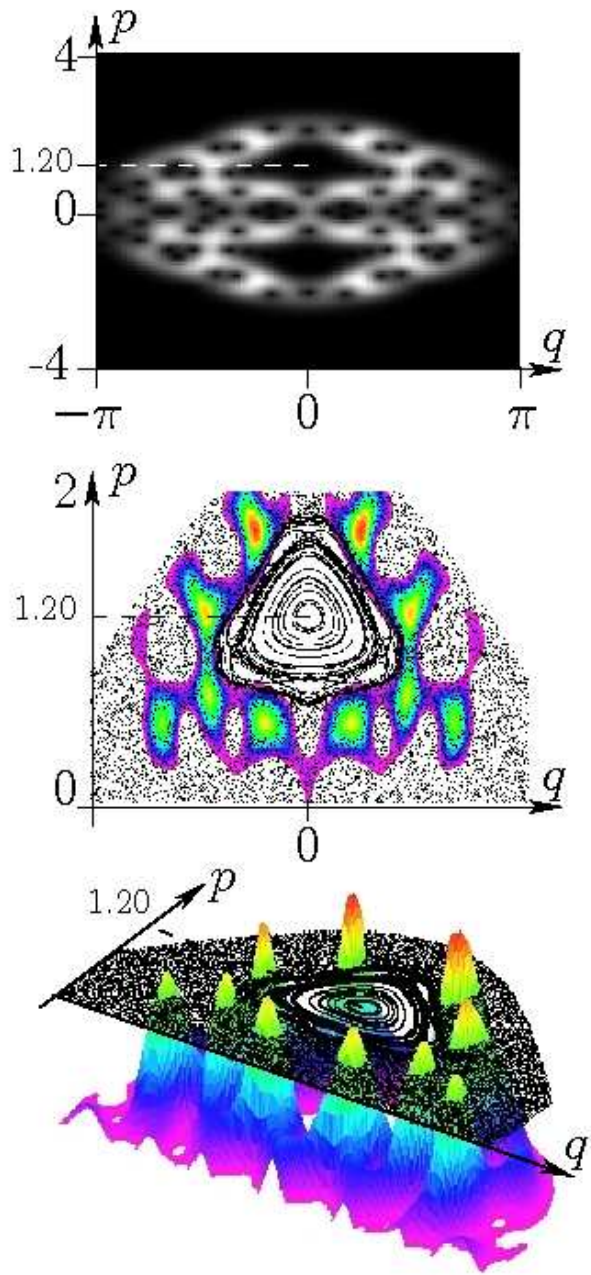}
\caption{\label{fig:janus}A Janus state: it is strongly coupled to the
most excited resonant state in the stable island and to the chaotic
states. In the random matrix model, it is ``seen'' as chaotic from the
regular states and almost ``regular'' from the chaotic sea. Here we
plot several views of the Husimi distribution of a Janus state and
compare its localisation with the classical phase-space structures
($\gamma=0.72$,
$\hbar^{-1}=30$, 
and $\epsilon=-0.007774$).}
\end{figure}

The analysis given so far, however, shows that 
the tunnelling enhancement cannot
be explained with just one crossing of the $\epsilon_0^\pm$ doublet
by~$\epsilon_m^\pm$. This global crossing is actually made up of many
elementary crossings whose contributions cannot be individually
distinguished. In effect, this game involves many players among which 
are the states that are completely delocalised in the chaotic sea.  
There are also states 
(``beach'' states \cite{Doron/Frischat95a,Frischat/Doron98a} or ``Janus''
states) that are strongly coupled on one
side to the regular states and on the other side to the chaotic ones.
It is scarcely surprising that these states have their Husimi
distribution localised near the borderline between the stable islands
and the chaotic sea (see FIG.~\ref{fig:janus}).

\section{Resonance-assisted tunnelling}\label{sec:rat}

\begin{figure}[!ht]
\center
\includegraphics[width=9cm]{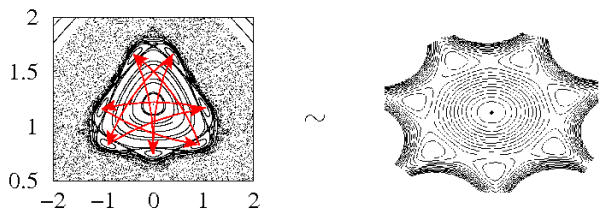}
\caption{\label{fig:internalstructure}
Left panel: classical phase space in the vicinity of the regular island.
The straight solid lines indicate in which order the sub-islands are
``visited'' in the course of time evolution.
The right panel shows the phase space structure that would result from the
pendulum-like Hamiltonian \eqref{eq:heff} describing the dynamics in the
vicinity of the 3/7 resonance.}
\end{figure}

The above discussion has provided overwhelming evidence for the
relevance of nonlinear resonances in the dynamical tunnelling process.
We now focus on the quantitative evaluation of the influence of such
resonances, which was presented in detail in
Refs.~\cite{Brodier+01a,Brodier+02a,Eltschka/Schlagheck05a,Schlagheck+05a}.
To this end, we formally introduce an integrable Hamiltonian that
approximately reproduces the dynamics within the regular island.  For
the upper island at $p \simeq 1$, such an integrable system can be
explicitly obtained by leaving out the $\gamma_+$-dependent term in
the Hamiltonian (\ref{eq:ham}).  Performing the time-dependent
canonical transformation $q \mapsto \tilde{q} = q - t$ to the frame
that co-propagates with the resonant orbit, this integrable
Hamiltonian reads
\begin{equation}
  H_0(p,\tilde{q}) = \frac{(p-1)^2}{2} - \frac{\gamma_-}{2} \ \cos(\tilde{q}) \, .
  \label{eq:h0}
\end{equation}
Canonical perturbation theory \cite{Lichtenberg/Lieberman83a} can be applied
on the basis of Eq.~(\ref{eq:h0}) in order to obtain an improved integrable
description that is in good agreement with the motion in the regular island
also at finite values of $\gamma$.

In the dynamics generated by $H_0$, the regular island is embedded in
a phase-space domain of bounded elliptic motion in momentum space.
Within this bounded domain, action-angle variables $(I,\theta)$ can be
introduced, which respectively correspond to the area enclosed by an
elliptic invariant orbit as well as to the propagation time that
elapses along this orbit.  In this action-angle variable
representation, we have $H_0(p,\tilde{q}) \equiv H_0(I)$, and the full
time-dependent Hamiltonian (\ref{eq:ham}) is formally written as
$H(I,\theta,t) = H_0(I) + V(I,\theta,t)$ where $V$ represents a weak
perturbation within the island.

We now assume the presence of a prominent 
$s/\ell$ resonance within the regular island, where $s$ oscillations
around the island's center match $\ell$ periods of the driving.  This
resonance condition is satisfied at $\ell \omega_0(I) = s$ where
$\omega_0 \equiv dH_0/dI$ is the oscillation frequency along the
bounded orbit with action variable $I$.  The dynamics in the vicinity
of such a resonance can be approximately described by the
pendulum-like integrable Hamiltonian
\begin{equation}
  H_\mathrm{eff}(I,\vartheta)=\frac{(I-I_0)^2}{2m_0}+2V_0\cos(\ell\vartheta) \label{eq:heff}
\end{equation}
which is derived from $H(I,\theta,t)$ using secular perturbation theory
\cite{Lichtenberg/Lieberman83a}.
Here, $\vartheta = \theta - (s/\ell) t$ is the (slowly varying) angle variable that 
co-rotates with the resonance.
The resulting phase space structure of $H_\mathrm{eff}$ is plotted in 
Fig.~\ref{fig:internalstructure}.

Following the lines of Ref.~\cite{Ozorio84a}, we now evaluate the
influence of such a resonance in the corresponding quantum system by a
direct quantization of the effective pendulum Hamiltonian
(\ref{eq:heff}) in the modified angle variable $\vartheta$.  Apart
from a phase factor containing the Maslov index, the unperturbed
eigenstates of $H_0$ are then given by the plane waves $\langle
\vartheta | n \rangle = \exp( i n \vartheta )$.  In the co-rotating
frame, their eigenenergies approximately read
\begin{equation} 
  E_n = \frac{[\hbar(n+1/2) - I_0]^2}{2m_0} \label{eq:en}
\end{equation}
using the fact that $I_n = \hbar(n+1/2)$ are the quantized action
variables within a regular island of elliptic shape.  The
$\vartheta$-dependent term in Eq.~(\ref{eq:heff}) introduces couplings
between the states $|n\rangle$ and $|n \pm \ell\rangle$ with the
coupling matrix element $V_0$.  These couplings give rise to
perturbative chains by means of which eigenstates with low and high
excitations within the bounded domain are connected to each other.

The pendulum Hamiltonian (\ref{eq:heff}) can be considered to be
appropriate for $I < I_c$ where $I_c$ denotes the action variable of
the outermost invariant elliptic curve of the regular island.  The
regime beyond this ``chaos border'' is characterized by the presence
of multiple overlapping resonances, which implies that the unperturbed
states with $I_n > I_c$ can be assumed to be strongly coupled to each
other by many different matrix elements.  Such couplings would also
occur between ``bound'' and ``unbound'' eigenstates of the integrable
Hamiltonian $H_0$ (which are, respectively, located within and outside
the domain of bounded motion that embeds the regular island) as well
as between states that are located in the vicinity of the two
different islands at $p \simeq 1$ and $p \simeq -1$.  In this way, an
efficient two-step mechanism is introduced by which two
symmetry-related quasimodes that are localized in the upper and lower
island, respectively, are coupled to each other: the nonlinear
resonance connects those quasimodes to the states in the chaotic
domain, and the latter ``see'' each other via strong matrix elements
of the full (Floquet) Hamiltonian of the driven system.

Fig.~\ref{fg:heff} schematically displays the effective Hamiltonian
matrix that governs this resonance- and chaos-assisted tunnelling
process between the two ``central states'' of the islands (i.e., given
by $|n=0\rangle$ in the above notation).  The matrix is restricted to
basis states of one particular symmetry class --- i.e., to states that
are ``even'' or ``odd'' with respect to time-reversal symmetry --- and
includes, for the sake of clarity, only those regular components to
which the central state is perturbatively connected via the $s/\ell$
resonance.
Altogether, $k_c$ regular states are included in the matrix, where the
integer $k_c$ is defined such that $I_{(k_c - 1)\ell} < I_c < I_{k_c \ell}$
holds true, i.e., $|(k_c - 1)\ell\rangle$ is still located in the island,
whereas $|k_c \ell\rangle$ is dissolved in the chaotic part.

We can now prediagonalize the upper left ``regular'' block of the Hamiltonian.
This yields, in lowest nonvanishing order in the perturbation strength
$|V_0|$ (which is much smaller than all relevant energy differences), the
modified central state as
\begin{equation}
  |\tilde{0}\rangle = |0\rangle + \sum_{k =1}^{k_c-1} \left( \prod_{k' = 1}^{k}
    \frac{V_0}{E_0 - E_{k' \ell}} \right) |k \ell\rangle \, .
  \label{eq:cmod}
\end{equation}
This perturbed central now exhibits a nonvanishing matrix element with one of
the states that are contained within the chaotic domain:
we obtain
\begin{equation}
  V_{\rm eff} \equiv \langle k_c\ell|H|\tilde{0}\rangle = V_0 \prod_{k=1}^{k_c - 1}
  \frac{V_0}{E_0 - E_{k\ell}} \label{eq:veff}
\end{equation}
which can be interpreted as the effective resonance-induced matrix element
between the central state and the chaotic domain.
In this way, our effective Hamiltonian matrix can be related to the
phenomenological matrix models that constitute the starting point of
the statistical theory of chaos-assisted tunnelling
\cite{Bohigas+93b,Tomsovic/Ullmo94a,Leyvraz/Ullmo96a}.

\begin{figure}[t]
  \begin{displaymath}
    \newlength{\Vrs}
    \settowidth{\Vrs}{$V_0$}
    \newlength{\csize}
    \setlength{\csize}{9.5em}
    H = \left(
      \begin{array}{ccccl}
        E_0 & V_0 & \phantom{\ddots}\\
        V_0 & E_\ell & V_0 & \phantom{\ddots}\\
        & V_0 & \ddots & \ddots \\
        & & \ddots & E_{(k_c-1)\ell} \hspace*{-0.3cm} 
        & \hspace*{0.2cm} V_0 \\[2mm]
        & & & \parbox[c][\csize][t]{\Vrs}{$V_0$} &
        \framebox{\parbox[c][\csize][c]{\csize}{\centering\large chaos}}
      \end{array}
    \right)
  \end{displaymath}
  \caption{Sketch of the effective Hamiltonian matrix that describes
    the coupling between the regular island and the chaotic domain for
    one particular symmetry class (i.e., for ``even'' or ``odd''
    states with respect to time-reversal symmetry).  The regular part
    (upper left band) includes only components that are coupled to the
    island's central state by the $s / \ell$ resonance.  In the simplest
    possible approximation, the chaotic part consists of a full
    sub-block with equally strong couplings between all basis states
    with actions beyond the outermost invariant torus of the island.
    \label{fg:heff}
  }
\end{figure}

In the simplest possible approach, we assume that the ``chaos block'' is
essentially homogeneous and can be well modeled by a random matrix from the
Gaussian Orthogonal Ensemble (GOE).
This assumption does not account for the influence of prominent partial
barriers to the classical transport, which can arise from broken invariant
tori (so-called ``Cantori'' \cite{MacKay+84a}) as well as from unstable
periodic orbits in the chaotic sea, and which would lead to an effective
division of the chaotic Hamiltonian into several sub-blocks that are weakly
connected to each other \cite{Bohigas+93b}.
Neglecting those partial barriers and performing the random matrix average over
the eigenvectors and eigenvalues of the chaos block gives rise to the Cauchy
distribution
\begin{equation}
  P(\Delta E_0) = \frac{2}{\pi} \frac{\overline{\Delta E_0}}
  {(\Delta E_0)^2 + (\overline{\Delta E_0})^2} \label{eq:cauchy}
\end{equation}
for the level splitting between the ``even'' and ``odd'' combination of 
central
states that are associated with the pair of regular islands
\cite{Tomsovic/Ullmo94a,Leyvraz/Ullmo96a}.
This distribution is characterized by the scale
\begin{equation}
  \overline{\Delta E_0} = \frac{2\pi V_{\rm eff}^2}{N_c \Delta_c}
\end{equation}
which contains the most dominant effective matrix element (\ref{eq:veff})
between the island's 
central 
 state and one of the chaotic states, the total
number $N_c$ of chaotic states, as well as the mean level spacing $\Delta_c$ in the
chaos block.
In our case of a periodically driven system, the latter is given by 
$\Delta_c = \hbar / N_c$, due to the fact that the chaotic states are, in the framework
of the Floquet approach, 
uniformly distributed at random
 in an energy window of the size 
$2 \pi \hbar / \tau = \hbar$ (we recall that the period of the driving equals $\tau = 2 \pi$).
As a consequence, we obtain $\overline{\Delta E_0} = 2\pi V_{\rm eff}^2/\hbar$.

The distribution (\ref{eq:cauchy}) is, strictly speaking, valid only for 
$\Delta E_0 \ll V_{\rm eff}$ and exhibits a cutoff at $\Delta E_0 \sim 2 V_{\rm eff}$
\cite{Leyvraz/Ullmo96a}, which ensures that the 
statistical expectation value $\langle\Delta E_0\rangle = \int_0^\infty x P(x) dx$ does not diverge.
However, since tunnelling rates and their parametric variations are typically
studied on a logarithmic scale (see Fig.~\ref{fig:comparison}), we compute
from Eq.~(\ref{eq:cauchy}) not the mean value $\langle\Delta E_0\rangle$, but rather the
average of the {\em logarithm} of $\Delta E_0$.
Our ``average'' level splitting $\langle\Delta E_0\rangle_g$ is therefore defined by the
{\em geometric} mean
$
  \langle\Delta E_0\rangle_g \equiv \exp \left[ \left\langle \ln(\Delta E_0) \right\rangle \right]
$
the evaluation of which does not involve the above cutoff; we obtain
the expression
\begin{equation}
  \langle\Delta E_0\rangle_g = \overline{\Delta E_0} = \frac{2\pi V_{\rm eff}^2}{\hbar} \, . 
  \label{eq:split}
\end{equation}
which, notably, is free of any adjustable parameter.
Hence, up to a trivial prefactor, the mean value of splittings is, in a
logarithmic-scale representation, given by the square of the coupling matrix
element (\ref{eq:veff}) between the island's 
central
 state and the chaos
\cite{Eltschka/Schlagheck05a,Schlagheck+05a}.

\begin{figure}[!ht]
\center
\includegraphics[width=8.5cm]{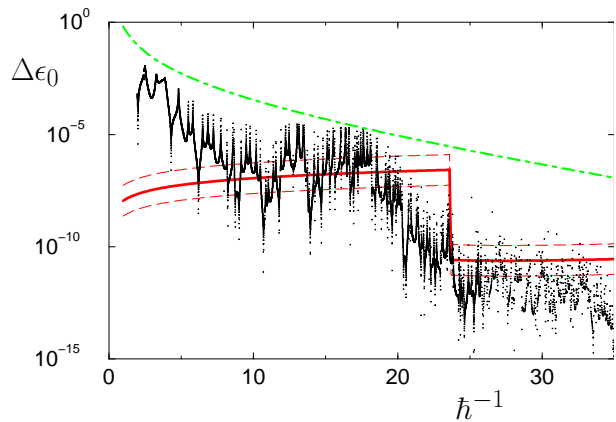}
\caption{\label{fig:splittings} Comparison between the numerically
calculated splittings of the driven pendulum at $\gamma=0.72$ (black
dots) and the semiclassical prediction according to
Eq.~(\ref{eq:split})
 (red solid line).  The semiclassical theory,
which is based on the prominent internal $3/7$ resonance within the
regular island, reproduces quite well the positions and heights of the
two plateaus that appear in the exact quantum splittings.  The dashed
lines indicate the size of the logarithmic standard deviation
according to Eq.~(\ref{eq:var}).  The green dot-dashed line displays
the prediction that would be obtained from Eq.~\eqref{eq:PNas} with 
$\Omega = 1$.
}
\end{figure}

Figure \ref{fig:splittings} shows the comparison with the exact
quantum splittings at $\gamma=0.72$, calculated by the numerical
diagonalization of the Floquet matrix.  The semiclassical prediction
(\ref{eq:split}) (solid line in Fig.~\ref{fig:splittings}) was
evaluated on the basis of the prominent 3/7 resonance, for which the
relevant parameters $I_0$, $m_0$ and $V_0$ that enter into the
pendulum Hamiltonian (\ref{eq:heff}) were entirely determined from
classical dynamics of the system: As in
Ref.~\cite{Eltschka/Schlagheck05a}, we compute for this purpose the
trace of the monodromy matrix associated with a stable or unstable
periodic point of the 3/7 resonance, as well as the phase space areas
that are enclosed by the inner and outer separatrices of the
resonance.  Indeed, those quantities remain invariant under the
canonical transformation to the action-angle variables
$(I,\vartheta)$, which means that the latter need not be explicitly
evaluated in order to obtain the effective coupling matrix element
$V_{\rm eff}$.

We see that the semiclassical theory reproduces quite well the two
plateaus that arise in the quantum splittings.  The drop in the
semiclassical splittings at $1/ \hbar \simeq 24$ occurs due to the
fact that the island supports more than seven locally quantized
eigenstates beyond this critical value of $1 / \hbar$; hence, two
perturbative steps instead of one are required in order to connect the
central state of the island to the chaotic domain.  This drop is
considerably softened in the quantum splittings, but can still be
identified, which confirms the relevance of the resonance-assisted
coupling mechanism in this tunnelling process.
Since nonlinear resonances represent a general feature of nonintegrable
Hamiltonian systems, we expect that the appearance of such step-like
sequences of plateaus is a \emph{generic} phenomenon, which is not
restricted to one particular system, but arises in various chaos-assisted
tunnelling processes.
This expectation is indeed confirmed by previous studies on the kicked Harper
\cite{Roncaglia+94a}, on the decay of nondispersive wave packets in driven
hydrogen \cite{Hornberger/Buchleitner98a}, as well as on the dynamical
tunnelling process of cold atoms \cite{Mouchet+01a,footnote4}, where
significant plateau structures were encountered in numerically calculated
tunnelling rates.

In addition to the mean value for the splittings, the probability distribution
(\ref{eq:cauchy}) can also be used in order to calculate the expectation value
for their \emph{logarithmic variance} characterizing the average size of
fluctuations on a logarithmic scale;
we obtain
\begin{equation}
  \langle \left( \ln \Delta E_0 - \ln \overline{\Delta E_0} \right)^2 \rangle = \pi^2 / 4 \label{eq:var} .
\end{equation}
This result is \emph{universal} in the sense that it does not depend on
system-specific parameters nor on $\hbar$.
The size of the corresponding standard deviation is indicated by the dashed
lines in Fig.~\ref{fig:splittings}, which are generated by multiplying
$\overline{\Delta E_0}$ with $e^{\pm \pi/2}$.
Clearly, the ``window'' defined by those dashed lines is exceeded near local
avoided crossings with chaotic states, where a large enhancement as well as a
complete suppression of the splittings may be induced \cite{Grossmann+91a}.
Apart from those exceptional events, however, the scale of the average
fluctuations of the splittings is well described by Eq.~(\ref{eq:var}).
The green dot-dashed line in Fig.~\ref{fig:splittings} displays the
prediction for the splittings that would be obtained from
Eq.~\eqref{eq:PNestimate} according to Ref.~\cite{Podolskiy/Narimanov03a}.
Here, the asymptotic expression
\begin{equation}\label{eq:PNas}
  \Delta\epsilon \simeq \frac{\hbar \Omega}{16 \pi N^3} \EXP{-2 (1-\ln 2) N}
\end{equation}
of Eq.~\eqref{eq:PNestimate}, valid for $N \equiv A / (2 \pi \hbar) \gg 1$, was
evaluated with the numerically calculated size $A$ of the island.
Being an entirely classical ($\hbar$-independent) quantity, the
system-dependent prefactor $\Omega$ was set equal to unity.
We clearly note a substantial disagreement between this semiclassical
estimation and the quantum splittings.

The theory of resonance-assisted tunnelling does not reproduce the
splittings in the deep quantum regime of larges values of $\hbar$,
where a different mechanism, possibly in the spirit of
Ref.~\cite{Podolskiy/Narimanov03a}, might induce the transition to the
chaos.  Moreover, the critical value of $1/ \hbar$ at which the drop
of the splittings from the first to the second plateau occurs is
significantly overestimated by our semiclassical approach.  This could
be due to the presence of \emph{partial barriers} in the chaotic phase
space domain, such as ``Cantori'' \cite{MacKay+84a}, which are known
to inhibit the quantum flux at not too small values for $\hbar$
\cite{Geisel+86a,Maitra/Heller00a}.  If such a Cantorus is manifested
in the vicinity of the regular island, the effective ``quantum'' size
of the island could be considerably enhanced as compared to $A$, which
would reduce the value of $1/ \hbar$ at which exactly $\ell$ quantum
states are localized around the island.  This observation is indeed in
accordance with the manifestation of Janus states
\cite{Doron/Frischat95a,Frischat/Doron98a} in the spectral analysis
(see Section \ref{sec:multilevels}).

Significant deviations of the splittings from the semiclassical prediction 
\eqref{eq:split} are also to be expected in the deep semiclassical regime
where a multitude of steps ($k \gg 1$) would be needed to connect the central
state to the chaotic domain according to the expression \eqref{eq:veff}.
In this regime, the coupling via the $s / \ell$ resonance --- which also
represents a dynamical tunnelling process as was pointed out in
Ref.~\cite{Brodier+02a} --- can again be assisted by the presence of another
nonlinear $s' / \ell'$ resonance, as long as this is permitted by the associated
selection rule ($n \mapsto n + \ell'$).
Such a $s' / \ell'$ resonance would generally exhibit a lower effective coupling
strength $V_0'$ and is typically of higher order than the $s / \ell$ resonance
(i.e., $\ell'$ and $s'$ are typically larger than $\ell$ and $s$, respectively).
The relevance of this multi-resonance coupling mechanism was demonstrated in
the near-integrable kicked Harper model where the semiclassical tunnelling
process involves a sequence of three nonlinear resonances
\cite{Brodier+01a,Brodier+02a,Schlagheck+05a}.

\begin{figure}[!ht]
\center
\includegraphics[width=6cm]{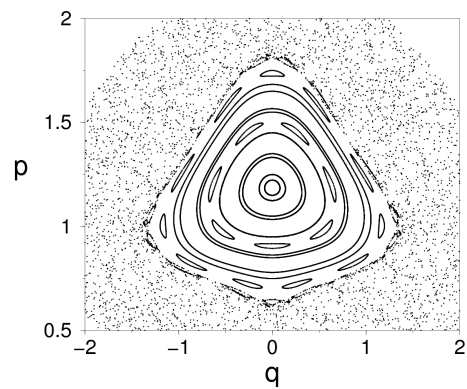}
\caption{\label{fig:ps67}
Classical phase space at $\gamma = 0.67$ in the vicinity of the regular island
$\mathcal{I}_+$.
The two chains of sub-islands correspond to the $3/7$ as well as to the $5/11$
resonance (inner and outer chain, respectively).}
\end{figure}

In the case of the driven pendulum at $\gamma = 0.72$, such multi-resonance
processes cannot be observed within the range of values for $1 / \hbar$ at which
precise quantum calculations of the splittings can be performed.
This is different, however, for $\gamma = 0.67$.
At this value of the coupling parameter, the regular island $\mathcal{I}_\pm$
exhibits two nonlinear resonances of almost equal importance:
a $3/7$ resonance, located closer to the center of the island than at 
$\gamma = 0.72$, and a $5/11$ resonance, located close to the chaos border 
(see Fig.~\ref{fig:ps67}).
This implies that a two-step tunnelling process involving both resonances can
be encountered at finite values for $1 / \hbar$.
Assuming that each resonance contributes with one single perturbative step
(which is the case at the values of $1 / \hbar$ considered here), the
corresponding matrix element that connects the central state to the
chaotic domain would read
\begin{equation}
  V_\mathrm{eff} = \frac{V_0^{(3/7)}}{E_0^{(3/7)} - E_7^{(3/7)}} V_0^{(5/11)}
  \, . \label{eq:veff2}
\end{equation}
Here, $V_0^{(s/ \ell)}$ represents the pendulum coupling strength of the
effective Hamiltonian \eqref{eq:heff}, and $E_n^{(s/ \ell)}$ denote the energies
\eqref{eq:en} in the co-rotating frame that is defined with respect to the 
$s/ \ell$ resonance.
This expression requires that the $7$th excited state, i.e.\ the first state
to which the central state of the island is coupled via the $3/7$ resonance,
is located \emph{in between} the two resonances, which implies that the action
variable $I_7$ of this state ought to be lower than the action variable
$I_0^{(5/11)}$ of the $5/11$ resonance.
This condition turns out to be valid for $1 / \hbar > 28.8$.

\begin{figure}[!ht]
\center
\includegraphics[width=8.5cm]{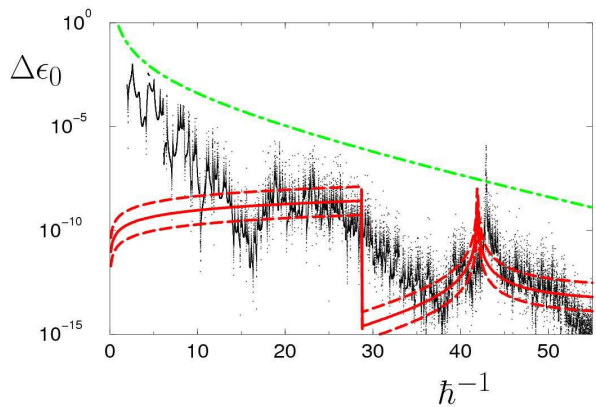}
\caption{\label{fig:splittings67}
Comparison between the numerically calculated splittings of the driven
pendulum at $\gamma=0.67$ (black dots) and the semiclassical prediction 
(red solid line).
The latter was evaluated by a single-resonance process via the $5/11$
resonance for $1 / \hbar < 28.8$, and by a two-resonance process according to
Eq.~\ref{eq:veff2} involving also the $3/7$ resonance for $1 / \hbar > 28.8$.
As in Fig.~\ref{fig:splittings}, the dashed lines indicate the size of the
logarithmic standard deviation according to Eq.~(\ref{eq:var}), and the green
dot-dashed line displays the prediction obtained from Eq.~\eqref{eq:PNas}.
}
\end{figure}

Fig.~\ref{fig:splittings67} displays the quantum splittings at $\gamma= 0.67$.
  The red solid line denotes the semiclassical prediction
\eqref{eq:split} for the splittings, which is for $1 / \hbar < 28.8$
calculated by the single-step process via the $5/11$ resonance (which
has the larger matrix element $V_0^{(s/ \ell)}$ and should therefore
dominate compared to the $3/7$ resonance) and for $1 / \hbar > 28.8$
obtained through the two-step process that is described by
Eq.~\eqref{eq:veff2}.  As in the case of $\gamma=0.72$, the overall
agreement between the semiclassical and the quantum splittings is
quite good, with significant deviations arising only in the deep
quantum regime at $1 / \hbar < 10$ as well as in the vicinity of the
cross-over between the single- and the two-step process, which is
artificially sharp in the semiclassical calculation.  Striking
evidence for the validity of the two-step process is the appearance of
a pronounced peak in the quantum splittings at $1 / \hbar \simeq 43$,
which arises due to a vanishing denominator in Eq.~\eqref{eq:veff2},
i.e., due to the fact the central state and the $7$th excited state
become near-degenerate in the co-rotating frame.  The position and
height of this peak are fairly well reproduced by the semiclassical
theory.

\section{Conclusion}

In summary, we provided clear evidence of the significance of nonlinear
resonances in the dynamical tunnelling process within the driven pendulum.
Indeed, the signature of the 3/7 resonance that is dominantly manifested in
the regular island at $\gamma=0.72$ is identified in various ways:
(a) The characteristic plateau structure in the level splittings sensitively
depends on whether or not the resonance is well developed within the regular
island,
(b) Floquet states that exhibit an appreciable overlap with the central Husimi
wavefunction of the island energetically correspond to the 7th excited
eigenstate within the island, and 
(c) a semiclassical expression for the tunnelling-induced splittings that is
based on the resonance-assisted coupling scheme is in good agreement with
the exact quantum data.
The validity of the resonance-assisted tunnelling mechanism is
furthermore confirmed by a striking peak in the quantum splittings at
$\gamma = 0.67$, which arises due to a two-step process involving two
different nonlinear resonances within the regular island.  Both in the
spectral analysis and in the semiclassical comparison, we identify
traces of Janus states that are located in the chaotic vicinity of the
islands.  This indicates that partial barriers in the chaotic domain
could still be relevant in our system at the values of $\hbar$ under
consideration.

The present approach is presumably not suited for a full-blown semiclassical
theory of mixed regular-chaotic tunnelling in terms of complexified orbits, for
which the ansatz of Refs.~\cite{Shudo/Ikeda95a,Shudo/Ikeda98a,Shudo+02a}
provides a more convenient framework.
Our findings, however, provides essential ingredients to the 
\emph{interpretation} of such semiclassical theories, in the sense that
``direct'' and ``resonance-assisted'' processes ought to be somehow
represented in relevant combinations of such complex orbits.
Moreover, specific quantitative predictions for tunnelling rates, using only
easily accessible quantities of the classical dynamics of the system, can be
made on the basis of our approach, provided $\hbar$ is small enough for the
most dominant resonance to become relevant (which is roughly the case if $\ell
/ 2$ states fit into the island).
We therefore believe that the principle of resonance-assisted couplings will
represent the relevant paradigm also in the context of more complex dynamical
tunnelling processes, e.g., within systems that have two or more degrees of
freedom.

\textbf{Acknowledgments:} 
It is a pleasure to thank Denis Ullmo for sharing with us
his extensive experience on tunnelling.
A.M.\ acknowledges 
the generous hospitality of Dominique Delande at Laboratoire Kastler-Brossel
and his relevant remarks after reading the first proof of this manuscript.
P.S.\ acknowledges support from the Bayerisch-Franz\"osisches Hochschulzentrum
(BFHZ) and from the Deutsche Forschungsgemeinschaft (DFG).


\begin{thebibliography}{10}
\bibitem{footnote}
For multidimensional non-separable but integrable systems, these tori are
  precisely the relevant classical structures that allow to generalise
  \textsc{jwkb} 1d-tunnelling~\cite{Creagh94a}.
\bibitem{footnote2}
The only difference is that a third resonance term must be added
  to~\eqref{eq:ham} in order to fulfil some experimental constraints on the
  lasers configuration. How this can be done is explained in
  Refs.~\cite{Mouchet+01a,Hensinger+01a,Steck+01a} but the present paper,
  devoted to theoretical considerations only, will not introduce this
  irrelevant third term.
\end{thebibliography}
\end{document}